\documentclass[aip,cha,reprint]{revtex4-2}

\usepackage{amsmath, amsfonts, amssymb, amsthm, mathtools}
\usepackage[utf8]{inputenc}
\usepackage[T1]{fontenc}
\usepackage{mathptmx}
\usepackage{hyperref}
\usepackage{graphicx}
\usepackage[caption=false]{subfig}
\usepackage{lipsum}
\usepackage{xcolor}
\usepackage{empheq}
\usepackage[normalem]{ulem}

\theoremstyle{remark}

\theoremstyle{remark}

\theoremstyle{theorem}

\theoremstyle{theorem}

\newcommand{\LT}[1]{\widetilde{#1}}

\newcommand{\MFPT}[1]{\langle \tau_{b}(#1)\rangle}

\begin{document}
	
\title{Optimal conditions for first passage of jump processes with resetting}

\author{Mattia Radice}
\email[Corresponding author: ]{mradice@pks.mpg.de}
\affiliation{Max Planck Institute for the Physics of Complex Systems, 01187 Dresden, Germany}
\author{Giampaolo Cristadoro}
\affiliation{Dipartimento di Matematica e Applicazioni, Università degli Studi Milano-Bicocca, 20126 Milan, Italy}
\author{Samudrajit Thapa}
\affiliation{Max Planck Institute for the Physics of Complex Systems, 01187 Dresden, Germany}

\begin{abstract}
We investigate the first passage time beyond a barrier located at $b\geq0$ of a random walk with independent and identically distributed jumps, starting from $x_0=0$. The walk is subject to stochastic resetting, meaning that after each step the evolution is restarted with fixed probability $r$. We consider a resetting protocol that is an intermediate situation between a random walk ($r=0$) and an uncorrelated sequence of jumps all starting from the origin ($r=1$), and derive a general condition for determining when restarting the process with $0<r<1$ is more efficient than restarting after each jump. If the mean first passage time of the process in absence of resetting is larger than that of the sequence of jumps, this condition is sufficient to establish the existence of an optimal $0<r^*<1$ that represents the best strategy, outperforming both $r=0$ and $r=1$. Our findings are discussed by considering two important examples of jump processes, for which we draw the phase diagram illustrating the regions of the parameter space where resetting with some $0<r^*<1$ is optimal.
\end{abstract}

\maketitle
\begin{quotation}
	What is the best strategy to optimize the first passage time of stochastic processes? This question is of fundamental interest in various fields, including biology, ecology, finance and climate science. One strategy that has been much studied in recent years is stochastic resetting, which consists of interrupting the dynamics and restarting the process until first passage. Here we consider stochastic resetting for jump processes, which are defined as a sum of independent and identically distributed (i.i.d.) random variables, useful for describing, e.g., the motion of run-and-tumble particles, such as bacteria, and Lévy flights, employed for instance to model efficient foraging behaviour of predators in the presence of sparse targets. We are interested in minimizing the mean number of jumps required to cross a barrier or threshold. Depending on the initial distance from the barrier and/or the presence of a bias, the optimal resetting strategy may be to not reset at all, or to reset after each jump. By combining analytics and numerics, we determine the conditions under which it is instead optimal to adopt a nontrivial strategy of letting the dynamics evolve for more than a single step before resetting. Our results are therefore potentially applicable to designing experiments with an optimal protocol for search processes. 
\end{quotation}

\section{Introduction}
The study of first passage events for stochastic processes is a fundamental problem in fluctuation theory, with a plethora of applications \cite{Metzler2014,Red}. For example, first passage problems are formulated in biology to describe diffusion-controlled chemical reactions \cite{Kotomin1996,Metzler2014} and in ecology to evaluate the efficiency of different search strategies \cite{BenLovMor-2011} implemented by animals or microorganisms searching for food \cite{Bell,Klafter1990}. They are important in finance to establish the optimal time to sell an asset \cite{ShiXuZho-2008,MajBou-2008} or to determine the optimal trading strategy in the presence of transaction costs \cite{DeLDerPot-2012}. First passage events also play a central role in extreme value statistics \cite{Bray-2000,HarGod-2019,VezBarBur-2024}, where they are associated to the study of maxima, records and record-breaking events \cite{MajZif-2008,Maj-2010,GodMajSch-2016,MajPalSch-2020,Sin-2022,PROA,AOPR,RegDolBen-2023}.

Common to all these situations is the question of how the process can be \emph{optimized}, and thus great effort is put into the study of optimization strategies. In this regard, one of the strategies that has been most investigated in recent years and still attracts much interest is stochastic resetting, which consists in restarting the process after random time intervals until the first passage event occurs \cite{EvaMaj-2011,EvaMaj-2011-II}. With respect to the first passage properties, the main advantage of resetting is the possibility to minimize the average first passage time by choosing the optimal value of the external control parameter (typically, the reset rate \cite{EvaMaj-2011}, but also different mechanisms have been studied \cite{EulMet-2016,NagGup-2016,PalKunEva-2016,PalReu-2017,CheSok-2018,RAD-2022,EliReu-2023}). Very recent results show that resetting is also effective for controlling the spatial aspects of the first passage event, such as the landing position of a random flyer when it overshoots a barrier for the first time \cite{RADCri-2024}, which is crucial to the evaluation of the search efficiency \cite{LomKorMet-2008,PalCheMet-2014}. This and other intriguing features have generated a significant amount of research on the topic, with resetting applied to a variety of different systems in areas such as biology \cite{ReuUrbKla-2014,RotReuUrb-2015,RolLisSanGri-2016}, quantum physics\cite{Nav-2018,PerCarMag-2021,MagCarPer-2022,SevVal-2023,YinBar-2023,YinWanBar-2024,YinWanTor-2024}, algorithm optimization \cite{MonZec-2002}, and economics \cite{StoSanKoc-2021,San-2022,StoJolPal-2022}.

However, the greatest interest has occurred within the statistical physics community. After the seminal work of Evans and Majumdar \cite{EvaMaj-2011,EvaMaj-2011-II}, who applied stochastic resetting to Brownian motion, several further studies have been proposed. An important line of research, motivated by the experimental realization of resetting systems \cite{BesBovPet-2020,TalPalSek-2020}, extended the original model by considering a return dynamic to the restart position \cite{PalKusReu-2019,PalKusReu-2019PRE,MasCamMen-2019PRE,MerBoyMaj-2020,PalKusReu-2020,BodSok-2020-BrResI,GupPlaKun-2021,SanDasNat-2021,RAD-2022}, or a time cost to perform the reset \cite{SunBlyEva-2023}. Another important line of research was aimed at applying resetting to different types of stochastic models, including the telegraph process and run-and-tumble motion \cite{EvaMaj-2018,Mas-2019,SanBasSab-2020,RAD-2021,SanKocMet-2022,SanIom-2024,GorSevCha-2024}, heterogeneous diffusion \cite{WanCheKan-2021,LenLenGui-2022,SanDomKoc-2022}, and Lévy flights \cite{KusMajSabSch-2014,KusGud-2015,StaWer-2021}. General results on the properties of processes with restart have been obtained in \cite{Reu-2016,PalReu-2017,Bel-2018,CheSok-2018,RayReu-2021,StaBel-2023,Bel-2024}. The effects of resetting have also been investigated in stochastic thermodynamics \cite{FucGolSei-2016,PalRah-2017,GupPlaPal-2020,MorOlsKri-2023,OlsGupMor-2024,OlsGup-2024,Sin-2024} and extreme value statistics \cite{MajPalSch-2020,SinPal-2021,GuoYanChe-2023,GuoYanChe-2024}. Recent lines of research focus on the impact of resetting on the ergodic properties of various systems \cite{WanCheKan-2021,WanCheMet-2022,VinCheWan-2022,BarFlaMen-2023}, its application to granular systems \cite{Bod-2024}, and in the context of search processes with mortal searchers \cite{RAD-2023,BoyMerMaj-2024} or with a variable number of searchers \cite{CamMen-2024}.

In this paper we follow the line aimed at investigating the effects of stochastic resetting on the first passage properties of random walks, which are described by a discrete time parameter \cite{KusMajSabSch-2014,BonPal-2021,RiaBoyHer-2020,BonPal-2021arXiv,FlyPil-2021,BonPalReu-2022,RiaBoyHer-2022,RAD-2022-Gill,DasGiu-2022}. We focus in particular on one-dimensional jump processes, which for us are a sum of i.i.d. continuos random variables $\eta_i$, such that the state of the process at time $n$ is $x_n=\eta_1+\dots+\eta_{n}$. Despite their simplicity, jump processes are important models for describing the dynamics of many systems and are subject to continuous research \cite{VezBarBur-2019,WanVezBur-2019,KliVoiBen-2022,VezBarBur-2024,BasVazBur-2024}.

The outline of the paper is as follows: in Sec. \ref{s:Mod} we define the model and motivate our work; in Sec. \ref{s:FPT} we present out theoretical findings, which are discussed for two particular classes of jump processes in Sec. \ref{s:Res}; finally in Sec. \ref{s:Conc} we draw our conclusions. 

\section{Model and Motivation}\label{s:Mod}
We consider a discrete-time random walk in one dimension that evolves according to
\begin{equation}\label{eq:dyn}
	x_{n}=\begin{cases}
		x_{n-1}+\eta_{n}&\text{with Prob. }1-r\\
		\eta_{n} &\text{with Prob. }r,
	\end{cases}
\end{equation}
where $\eta_n$ are independent and identically distributed random variables drawn from a common distribution, which we assume continuous with PDF $\lambda(\eta)$. The starting point of the walk is $x_0=0$, while $0\leq r\leq1$ is a constant. In practice, before each step we toss a biased coin and decide whether to perform the next jump from the current position or from the origin. Since the origin corresponds to the starting point, we can call $r$ the restart probability.

The model of restart that we are considering follows the same convention used by Bonomo and Pal \cite{BonPal-2021}. A very similar model was considered by Campos and Méndez \cite{CamMen-2015}, albeit in a continuous-time formalism. In other works instead \cite{KusMajSabSch-2014,RiaBoyHer-2020,BonPal-2021arXiv,FlyPil-2021,BonPalReu-2022,RiaBoyHer-2022} a different convention is chosen, where instead of Eq. \eqref{eq:dyn} the evolution is governed by
\begin{equation}\label{eq:dyn_orig}
	x_{n}=\begin{cases}
		x_{n-1}+\eta_{n}&\text{with Prob. }1-r\\
		x_0 &\text{with Prob. }r.
	\end{cases}
\end{equation}
The difference is that with probability $r$ the walker jumps \emph{to} the starting point, and not $\emph{from}$ the starting point as in Eq. \eqref{eq:dyn}. Therefore, for $r=1$ one has a degenerate case where the walker never leaves $x_0$. This has obvious implications on the first passage properties of the resetting process, which is the problem we want to study in this paper. Indeed, since the evolution freezes for $r=1$, one obtains a diverging first passage time. The choice of the mechanism defined in Eq. \eqref{eq:dyn} allows us to avoid this divergence. In this way, studying the stochastic model of Eq. \eqref{eq:dyn} with some $0<r<1$ corresponds to considering an intermediate situation between the reset-free process ($r=0$) and a sequence of i.i.d. jumps ($r=1$). The relevant question then becomes whether there is an intermediate situation that outperforms both the cases $r=0$ and $r=1$.

\section{First passage time}\label{s:FPT}
Suppose that a barrier or threshold is located at $b\geq0$. We are interested in $F_r(n;b)$, which is the probability that $x_n$ becomes larger than $b$ for the first time at step $n$, with the restart probability set to $r$. Let us denote with $Q_0(n;b)$ the \emph{survival probability} of the random walk in absence of resetting, namely, the probability that the process, with $r=0$, does not exceed $b$ up to time $n$. We can write a simple renewal equation \cite{KusMajSabSch-2014,BonPal-2021}:
\begin{align}
	F_r(n;b)=& r\sum_{m=0}^{n-1}(1-r)^{m-1}Q_0(m;b)F_r(n-m;b)\nonumber\\
	&+(1-r)^nF_0(n;b).
\end{align}
By introducing the generating functions
\begin{subequations}
	\begin{align}
		\LT{F}_r(z;b)&=\sum_{n=1}^{\infty}z^nF_r(n;b)\\
		\LT{Q}_0(z;b)&=\sum_{n=0}^{\infty}z^nQ_0(n;b),
	\end{align}
\end{subequations}
with $0<z<1$, we obtain
\begin{equation}
	\LT{F}_r(z;b)=\frac{\LT{F}_0(z_r;b)}{1-r\LT{Q}_0(z_r;b)},\quad z_r=(1-r)z,
\end{equation}
and by using $\LT{F}_0(z;b)=1-(1-z)\LT{Q}_0(z;b)$ (see e.g. Feller \cite{Fell-II}), we finally get
\begin{equation}
	\LT{F}_r(z;b)=\frac{1-(1-z_r)\LT{Q}_0(z_r;b)}{1-r\LT{Q}_0(z_r;b)}.
\end{equation}
As a measure of the speed of the process, we consider the mean first passage time (MFPT), which is given by
\begin{align}
	\MFPT{r}&=\left.\frac{\partial\LT{F}_r(z;b)}{\partial z}\right\vert_{z=1}\nonumber\\
	&=\frac{(1-r)\LT{Q}_0(1-r;b)}{1-r\LT{Q}_0(1-r;b)}.\label{eq:MFPT}
\end{align}
We note that this is a different formulation of Eq. (20) in Ref. \cite{BonPal-2021}. We can see now that by setting $r=1$ we obtain
\begin{equation}\label{eq:MFPT_r1}
	\MFPT{1}=\frac{1}{1-Q_0(1;b)},
\end{equation}
which is indeed finite if $Q_0(1;b)<1$. In other words, in contrast to what happens by using Eq. \eqref{eq:dyn_orig}, $\MFPT{1}$ does not diverge as long as the process can be completed in a single step.

\subsection{Optimization of the MFPT}
In the optimization problem we ask if there is a \emph{nontrivial} $0<r^*<1$ that minimizes the MFPT. We call instead trivial the situations where the minimum corresponds to $r^*=0$ or $r^*=1$. To study this problem it is thus also relevant the value of $\MFPT{0}$:

\begin{itemize}
	\item[i)] Suppose that $\MFPT{0}<\MFPT{1}$, i.e., a random walk is faster, on average, than a sequence of i.i.d. jumps. As we will see later, this can happen, e.g., if there is a bias towards the barrier. Then a sufficient condition to have a nontrivial $r^*$ is that the derivative $d\MFPT{r}/dr$ be negative for $r\to0^+$. This condition has been thoroughly studied in the literature \cite{BonPal-2021,BonPal-2021arXiv,FlyPil-2021,BonPalReu-2022,RiaBoyHer-2022}, and in our model it corresponds to the criterium derived in \cite{BonPal-2021}:
	\begin{equation}\label{eq:Cond_Pal}
		\frac{\langle\tau_b(0)^2\rangle-\MFPT{0}^2}{\MFPT{0}^2}>1-\frac{1}{\MFPT{0}}.
	\end{equation}
	
	\item[ii)] Suppose now that $\MFPT{0}>\MFPT{1}$, which means that a random walk is, on average, slower than a sequence of jumps. Note that this is the typical situation one obtains for symmetric jumps, for which, as a consequence of a theorem of Sparre-Andersen \cite{Sparre}, the MFPT of the random walk diverges. In this case, a sufficient condition to have a nontrivial $r^*$ is that the derivative $d\MFPT{r}/d r$ be positive for $r\to1^-$. By expanding $\MFPT{r}$ close to $r=1$ we find
	\begin{equation}
		\MFPT{r}\sim\frac{1}{1-Q_0(1;b)}
		+(1-r)\frac{Q_0(2;b)-Q_0(1;b)^2}{(1-Q_0(1;b))^2}
	\end{equation}
	hence the derivative is positive if
	\begin{equation}\label{eq:Cond}
		Q_0(2;b)<Q_0(1;b)^2.
	\end{equation}
\end{itemize}

We note that \eqref{eq:Cond} has a simple probabilistic interpretation in terms of conditional probability. Indeed, consider the space of all random walks of two steps, and call $\Omega$ the subset of all the random walks that occupy position $x_1\leq b$ after one step, and position $x_2\leq b$ after two steps. In other words, $\Omega$ is the intersection of the set of all random walks with $x_1\leq b$, and the set of all random walks with $x_2\leq b$, and its probability is thus by definition $Q_0(2;b)$. Since the probability of the intersection can be written in terms of the conditional probability, we have
\begin{equation}
	Q_0(2;b)=\Pr\lbrace x_2\leq b\,|\,x_1\leq b\rbrace\Pr\lbrace x_1\leq b\rbrace.
\end{equation}
Noting that $\Pr\lbrace x_1\leq b\rbrace=Q_0(1;b)$ and plugging this into Eq. \eqref{eq:Cond}, one obtains
\begin{equation}
	\Pr\lbrace x_2\leq b\,|\,x_1\leq b\rbrace<\Pr\lbrace x_1\leq b\rbrace.
\end{equation}
We immediately observe that this inequality can not be satisfied if $b=0$. Indeed, if $b=0$ a random walk can survive after the first step only if the first jump is performed in the negative direction, so that $x_1<0$. Then the conditional probability that this random walk remains below $b=0$ after the second jump is certainly larger than $Q_0(1;b=0)$, because now the walker can also perform a jump in the positive direction, as long as the size of this jump is smaller than $|x_1|$. On the contrary, if $b>0$ the survival probability after one step takes into account also the walks that perform a first jump of length smaller than $b$ in the positive direction. Since these walks are closer to the barrier, they contribute to decreasing the conditional probability and can make the inequality true. This means that to find situations where \eqref{eq:Cond} is satisfied, we need to study survival problems with strictly positive $b$. This is generally very hard to do analytically, but the advantage of Eq. \eqref{eq:Cond} is that it only involves properties of the reset-free random walk after two steps. Hence, in principle, it can always be studied numerically. 

\section {Results}\label{s:Res}

\subsection{Skewed Laplace distribution}
\begin{figure*}
	\includegraphics[width=\linewidth]{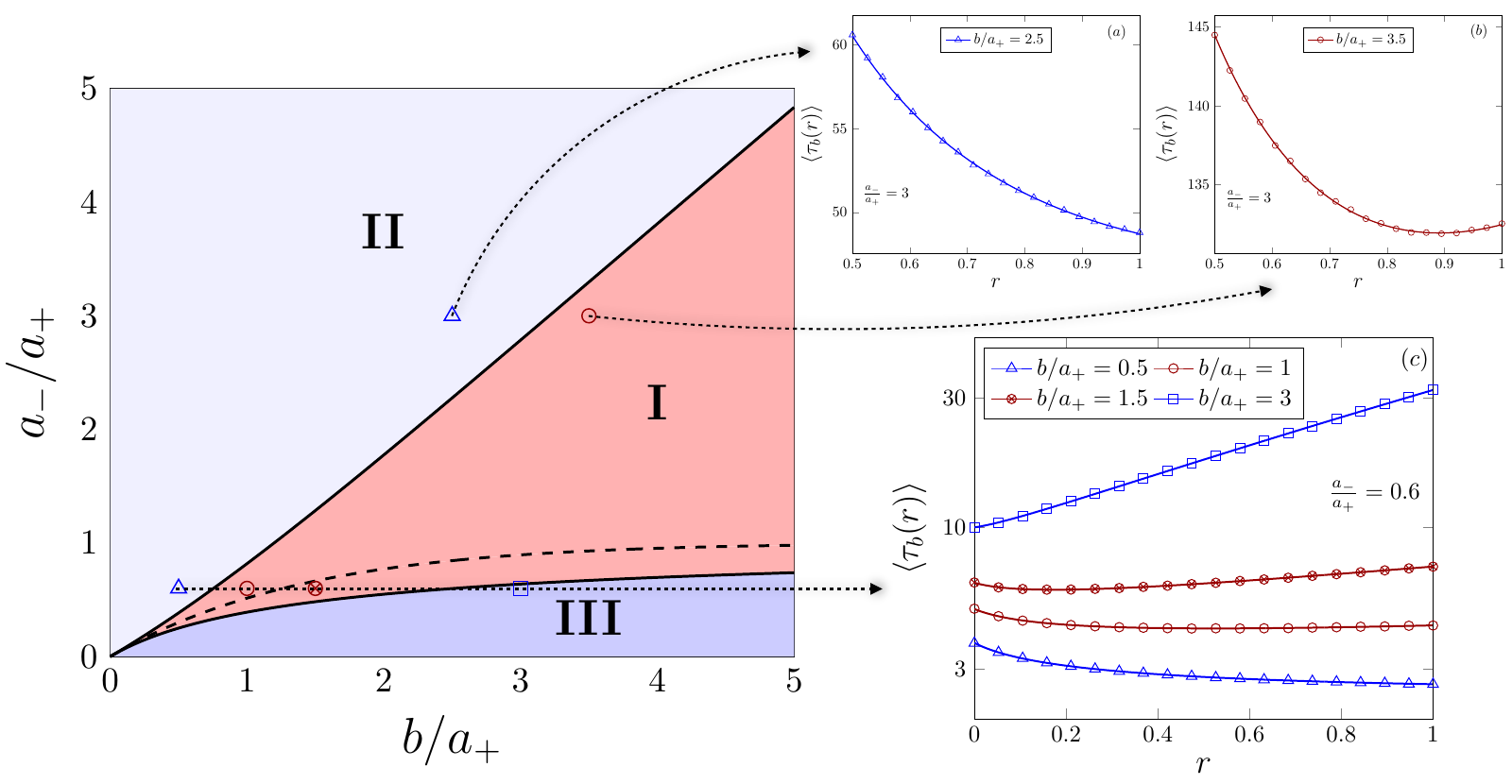}
	\caption{Phase diagram for random walks with jumps drawn from the skewed Laplace distribution, with $\lambda(\eta)$ given by Eq. \eqref{eq:PDF_exp_bias}. We set $b/a_+=X$ and $a_-/a_+=Y$. Region I is the domain where there exists a nontrivial $0<r^*<1$ that minimizes the $\MFPT{r}$. In region II $\MFPT{r}$ is a monotonic decreasing function of $r$, with global minimum at $r^*=1$. In region III instead we observe the opposite behaviour, and $\MFPT{r}$ is a monotonic increasing function of $r$, attaining its global minimum at $r^*=0$. Note that this region extends in the $Y$-direction only as long as $a_-/a_+<1$, i.e., when the jump distribution is biased toward the barrier. Region I is delimited by the critical curves $\overline{Y}_1(X)$ and $\overline{Y}_2(X)$ given by Eq. \eqref{eq:Y1} and \eqref{eq:Y3} respectively, and contains the dashed curve representing $Y(X)$ given by Eq. \eqref{eq:Y2}, which splits I into two subregions, according to whether $\MFPT{0}>\MFPT{1}$ (between $\overline{Y}_1(X)$ and $Y(X)$) or vice versa (between $Y(X)$ and $\overline{Y}_2(X)$). Panels (a), (b) and (c) show several examples of MFPT curves. The data are the results of numerical simulation and are compared to the exact curves given by Eq. \eqref{eq:BiasExp_MFPT_exact}, showing excellent agreement.}
	\label{fig:ExpBias_PD}
\end{figure*}
We first discuss a relevant example that allows us to obtain exact results and study the effects of a bias towards or away from the barrier. Let
\begin{equation}\label{eq:PDF_exp_bias}
	\lambda(\eta)=\frac{e^{\ell \eta/a^2}}{2\sqrt{\ell^2+a^2}}e^{-|\eta|\sqrt{\ell^2+a^2}/a^2},
\end{equation}
be the PDF of the jumps. The parameter $2\ell$ represents the mean jump, $\int_{-\infty}^{+\infty}\eta\lambda(\eta)d\eta=2\ell$, which in general is nonvanishing, while $a>0$ is a length scale, which is related to the variance $\sigma^2$ by $\sigma^2=2(a^2+4\ell^2)$. This is a generalization of the well-known symmetric Laplace distribution, with PDF $\lambda(\eta)=\exp(-|\eta|/a)/(2a)$, which corresponds to the case $\ell=0$. The Laplace distribution has great relevance for describing the motion of active matter, in particular run-and-tumble particles \cite{MorDouMajSch-2020,MorDouMajSch-2020PRE,MorMajViv-2024}. Hence the PDF in Eq. \eqref{eq:PDF_exp_bias} may be useful to describe more general situations where the motion is unbalanced and there is a tendency to move in a certain direction, depending on the sign of $\ell$.

To show how Eq. \eqref{eq:PDF_exp_bias} is obtained, let us first observe that a symmetric Laplace distributed jump is the sum of two exponentially distributed jumps, one positive and one negative, both with mean length $a$. Indeed, the characteristic function of a symmetric Laplace random variable $\hat{\lambda}(k)=1/(1+a^2k^2)$ can be factorized as $\hat{\lambda}(k)=\hat{\lambda}_+(k)\hat{\lambda}_-(k)$, where
\begin{equation}
	\hat{\lambda}_\pm(k)=\frac{1}{1\mp iak}.
\end{equation}
The two factors are precisely the characteristic functions of two exponentially distributed random variables, one positive and one negative, with mean $\pm a$, respectively, and $\hat{\lambda}(k)$ is thus the characteristic function of their sum. The generalization to the skewed Laplace case follows by considering
\begin{equation}\label{eq:CF_exp_bias}
	\hat{\lambda}(k)=\frac{1}{1-ia_+k}\cdot\frac{1}{1+ia_-k}=\frac{1}{1+a^2 k^2-2i\ell k},
\end{equation}
with $a_+$ and $a_-$ positive and distinct parameters. The rhs is indeed the Fourier transform of the PDF in Eq. \eqref{eq:PDF_exp_bias}, with $a^2=a_+a_-$ and $2\ell=a_+-a_-$. If $a_+>a_-$ there is a tendency to move in the positive direction and the opposite is observed if $a_+<a_-$. The parameter $2\ell=a_+-a_-$ can thus be interpreted as the bias of the jumps.

This choice of the PDF allows us to solve the problem exactly. The first step is to compute the generating function $\LT{Q}_0(z;b)$, which can be obtained with the Wiener-Hopf factorization technique: the crucial point is the factorization of the function
\begin{equation}
	\mathcal{P}(w,z)=\frac{1}{1-z\hat{\lambda}(w)},\quad w=k+is
\end{equation}
into the product of two functions $\mathcal{P}_\pm(w,z)$, with no singularities for $s>0$ and $s<0$ respectively \cite{Note}. By imposing that $\mathcal{P}_\pm(w,z)\to1$ as $|w|\to\infty$, the choice of the factor is unique. In general, determining $\mathcal{P}_\pm(w,z)$ explicitly is very difficult, but in the present case they can be deduced algebraically from $\mathcal{P}(w,z)$. Indeed, by using Eq. \eqref{eq:CF_exp_bias}, we can write
\begin{align}
	\mathcal{P}(w,z)&=\frac{1+a^2w^2-2i\ell w}{1-z+a^2w^2-2i\ell w}\\
	&=\underbrace{\frac{\theta_1(0)-iaw}{\theta_1(z)-iaw}}_{\mathcal{P}_+(w,z)}\cdot\underbrace{\frac{\theta_2(0)+iaw}{\theta_2(z)+iaw}}_{\mathcal{P}_-(w,z)},
\end{align}
where
\begin{align}
	\theta_1(z)&=\sqrt{\frac{\ell^2}{a^2}+1-z}-\frac{\ell}{a}\\
	\theta_2(z)&=\sqrt{\frac{\ell^2}{a^2}+1-z}+\frac{\ell}{a}.
\end{align}
Once we have the factors, the generating function $\LT{Q}_0(z;b)$ can be computed from the Laplace-inversion type integral \cite{RADCri-2024}
\begin{equation}
	\LT{Q}_0(z;b)=\frac{\mathcal{P}_-(0,z)}{2\pi i}\int_{\epsilon-i\infty}^{\epsilon+i\infty}\frac{e^{sb}}{s}\mathcal{P}_+(is,z)ds,
\end{equation}
with $\epsilon>0$, which yields
\begin{equation}\label{eq:Q_bias_exp}
	\LT{Q}_0(z;b)=\frac{1}{1-z}\left[1-\left(1-\frac{\theta_1(z)}{\theta_1(0)}\right)e^{-\theta_1(z)b/a}\right].
\end{equation}
We note that for $\ell=0$ we have $\theta_1(z)=\sqrt{1-z}$, and Eq. \eqref{eq:Q_bias_exp} indeed generalizes the result obtained for the symmetric Laplace distribution \cite{MajComZif-2005}. By plugging this into Eq. \eqref{eq:MFPT}, we obtain the exact expression of the MFPT:
\begin{equation}\label{eq:BiasExp_MFPT_exact}
	\MFPT{r}=(1-r)\frac{\theta_1(0)e^{\theta_1(1-r)b/a}+\theta_1(1-r)-\theta_1(0)}{r[\theta_1(0)-\theta_1(1-r)]}.
\end{equation}

We can now move on with the optimization problem. We recall that the condition for determining whether a nontrivial resetting strategy exists is different for the cases $\MFPT{0}<\MFPT{1}$ and $\MFPT{0}>\MFPT{1}$, see Sec. \ref{s:FPT}. Hence to proceed we first need to evaluate the behaviour of the MFPT as $r\to0^+$. From Eq. \eqref{eq:BiasExp_MFPT_exact}, we find that for $\ell\leq0$, i.e., if the bias is away from the target or neutral, the MFPT diverges as $r$ goes to zero. In particular:
\begin{equation}
	\MFPT{r}\sim
	\begin{cases}
		\frac1r\left(\frac{a_-}{a_+}e^{2|\ell|b/a^2}-1\right)&\text{if }\ell<0\\
		\frac1{\sqrt{r}}\left(1+\frac{b}{a_+}\right)&\text{if }\ell=0.
	\end{cases}
\end{equation}
The difference in the two asymptotic behaviours depends on the fact that for $\ell<0$ the walk in absence of resetting is transient (the probability of exceeding $b$ is smaller than one), while for $\ell=0$ is null-recurrent (the barrier is exceeded with probability one, but with a diverging MFPT). As observed in other cases, this leads to a difference in the order of the divergence for $r\to0^+$ \cite{RAD-2022-Gill}. On the other hand, for $\ell>0$ the small-$r$ behaviour is
\begin{equation}\label{eq:MFPT_bias_exp_r0}
	\MFPT{r}\sim\frac{b+a_+}{a_+-a_-}+\left[\frac{b^2/2}{(a_+-a_-)^2}-\frac{(b+a_+)a_-^2}{(a_+-a_-)^3}\right]r,
\end{equation}
hence the MFPT in absence of resetting is finite and equal to
\begin{equation}
	\MFPT{0}=\frac{b+a_+}{a_+-a_-}.
\end{equation}
Therefore, when $\ell>0$ this value must be compared with $\MFPT{1}$, which is given by
\begin{equation}
	\MFPT{1}=\left(1+\frac{a_-}{a_+}\right)e^{b/a_+},
\end{equation}
and to have $\MFPT{0}>\MFPT{1}$ the condition
\begin{equation}
	\frac{b+a_+}{a_+-a_-}>\left(1+\frac{a_-}{a_+}\right)e^{b/a_+}
\end{equation}
must be satisfied. This can be studied by introducing the positive variables $Y=a_-/a_+$ and $X=b/a_+$, so that the previous inequality can be recast as
\begin{equation}\label{eq:cond_for_Y}
	Y^2>1-e^{-X}(1+X).
\end{equation}
Note that $\ell=a_+(1-Y)$, hence the case $\ell>0$ corresponds to $0<Y<1$, and we need to study the inequality only in this range. We obtain that Eq. \eqref{eq:cond_for_Y} is satisfied for the points above the curve
\begin{equation}\label{eq:Y2}
	Y(X)=\sqrt{1-e^{-X}(1+X)}.
\end{equation}
By also noting that $\ell\leq0$ corresponds to $Y\geq1$, and recalling that in that case the MFPT in absence of resetting diverges, we conclude that the region of the $(X,Y)$ plane where $\MFPT{0}>\MFPT{1}$ is the region $X>0$, $Y>0$ above $Y(X)$.

When $\MFPT{0}>\MFPT{1}$ Eq. \eqref{eq:Cond} is sufficient to guarantee the existence of a nontrivial optimization. By computing the derivatives of $\LT{Q}_0(z;b)$ in Eq. \eqref{eq:Q_bias_exp}, we obtain the survival probabilities
\begin{align}
	Q_0(1;b)&=1-\frac{a_+e^{-b/a_+}}{a_++a_-}\\
	Q_0(2;b)&=1-\frac{a_+e^{-b/a_+}}{a_++a_-}\left(1+\frac{b}{a_++a_-}+\frac{a_+a_-}{(a_++a_-)^2}\right),
\end{align}
thus \eqref{eq:Cond} may be rewritten as
\begin{equation}
	\frac{a_+e^{-b/a_+}}{a_++a_-}>1-\frac{b}{a_++a_-}-\frac{a_+a_-}{(a_++a_-)^2}.
\end{equation}
Again, this inequality can be studied in terms of the positive variables $Y=a_-/a_+$ and $X=b/a_+$. In this way we can recast the condition as
\begin{equation}
	Y^2-h(X)Y-h(X)<0,
\end{equation}
where $h(x)=e^{-x}+x-1$ is a positive function for $x>0$. By defining the critical curve
\begin{equation}\label{eq:Y1}
	\overline{Y}_1(X)=\frac{h(X)+\sqrt{h(X)^2+4h(X)}}{2},
\end{equation}
the region of the $(X,Y)$ plane where the inequality is satisfied is the region $X>0$, $Y>0$ below this curve.

On the other hand, when $\MFPT{0}<\MFPT{1}$ a sufficient condition for nontrivial optimization is given by Eq. \eqref{eq:Cond_Pal}, which corresponds to requiring the derivative $d\MFPT{r}/dr$ to have a negative sign for $r$ close to zero. From Eq. \eqref{eq:MFPT_bias_exp_r0} we see that we must check
\begin{equation}
	\frac{b^2/2}{(a_+-a_-)^2}-\frac{(b+a_+)a_-^2}{(a_+-a_-)^3}<0.
\end{equation}
Again, we use $Y=a_-/a_+$ and $X=b/a_+$, so that this condition becomes
\begin{equation}
	\frac{X^2}{2(1-Y)^2}-\frac{(X+1)Y^2}{(1-Y)^3}<0,
\end{equation}
from which we obtain the curve
\begin{equation}\label{eq:Y3}
	\overline{Y}_2(X)=\frac{X}{4+4X}\left(\sqrt{X^2+8X+8}-X\right).
\end{equation}
The region of the $(X,Y)$ plane where the derivative of $\MFPT{r}$ has a negative sign for $r\to0^+$ is the area above this curve.

All these results are illustrated in Fig. \ref{fig:ExpBias_PD}, where the $(X,Y)$ plane is divided into three separate regions. Recall that $0<Y<1$ corresponds to a bias toward the barrier, $Y>1$ to a bias away from the barrier and $Y=1$ to the unbiased case. The interesting domain is region I, where a nontrivial optimization exists. This region is delimited by the critical curves $\overline{Y}_1(X)$ and $\overline{Y}_2(X)$, and is split into two subregions by $Y(X)$, according to whether $\MFPT{0}>\MFPT{1}$ or vice versa. In the remaining domains, $\MFPT{r}$ is a monotonic function of $r$: in region II the fastest strategy is resetting after each step ($r^*=1$), whereas in region III the situation without resetting is the fastest one $(r^*=0)$. Remarkably, the phase diagram shows that for any value of $Y$ there is a range of $X$ falling in region I. However, for $Y\geq1$, i.e., when the jumps are unbiased or biased away from the barrier, the range extends from a certain critical value of $X$, obtained from $\overline{Y}_1(X)$, to infinity. For $0<Y<1$ instead, the range is finite, delimited by the curves $\overline{Y}_1(X)$ and $\overline{Y}_2(X)$.

We can thus summarize as follows: if the bias is away from the target, or there is no bias, for small distances from the barrier the fastest strategy corresponds to resetting after each step. However, when the distance $b$ exceeds a certain critical value, the optimal strategy corresponds to a nontrivial resetting probability. If the walk has a tendency to move toward the barrier, the situation is initially similar, and the optimal strategy is nontrivial only after a critical value of $b$. However, when $b$ exceeds a second critical value, the optimal strategy becomes trivial again, but in this case the minimal MFPT is obtained for $r=0$.

\subsection{Lévy stable processes}
\begin{figure*}
	\includegraphics[width=\linewidth]{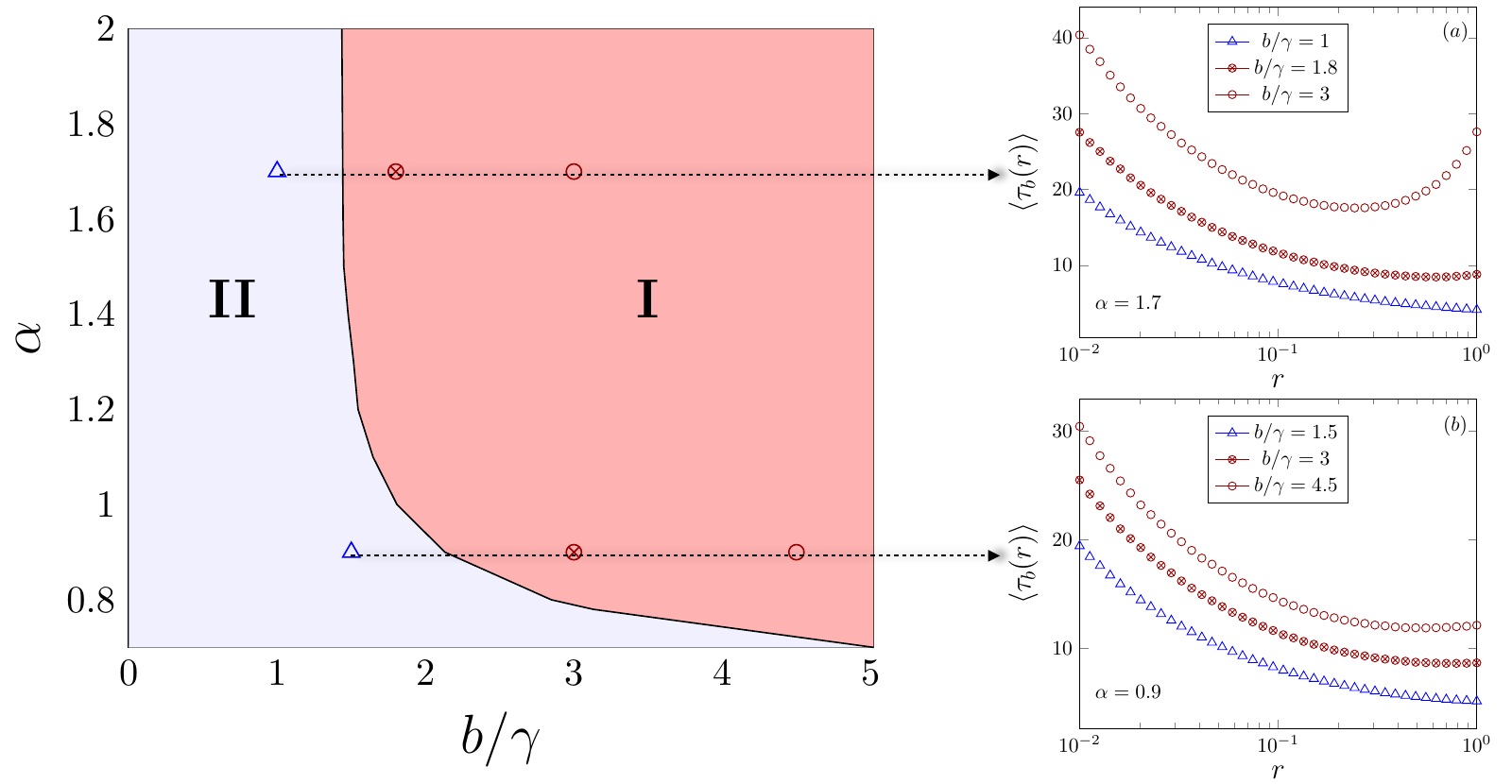}
	\caption{Phase diagram for Lévy flights with jumps drawn from Lévy stable laws, with $\hat{\lambda}(k)=e^{-|\gamma k|^\alpha}$. Region I is the domain where Eq. \eqref{eq:Cond} is verified, so $\MFPT{r}$ is guaranteed to have its global minimum for some $0<r^*<1$. In region II instead the existence of a nontrivial resetting probability is not guaranteed, and indeed in this domain our numerical simulations show a monotonic behaviour of  $\MFPT{r}$, suggesting that the minimum is always attained at $r^*=1$. Plots (a) and (b) show examples (with $\alpha=1.7$ and $\alpha=0.9$ respectively) of MFPT curves. The phase diagram does not show values $\alpha<0.7$ due to numerical difficulties in computing the critical point where region I begins, see text.}
	\label{fig:Levy_PD}
\end{figure*}
Lévy flights constitute a well-known class of jump processes with jumps drawn from a PDF $\lambda(\eta)$ characterized by an asymptotic power-law decay $|\eta|^{-1-\alpha}$, with $0<\alpha<2$. This means that the variance of the jump length diverges, as well as its mean for $0<\alpha\leq1$. Lévy flights are ubiquitous and find applications in animal foraging \cite{VisAfaBul-1996,Vis-2010}, human behaviour \cite{BroHufGei-2006,GonHidBar-2008}, and search processes \cite{LomKorMet-2008,BenLovMor-2011,PalCheMet-2014}, to cite a few.

Here we focus on the symmetric case, such that there is an even probability of jumping to right or left. This is the most studied case, even tough the asymmetric case has also generated interest in the literature \cite{PadCheDyb-2019,PadCheDyb-2020,PadSanKan-2022}. In this way, we are sure that the process in absence of resetting has a diverging MFPT and it is thus sufficient to evaluate the condition given in Eq. \eqref{eq:Cond}. It is important to remark that this condition depends on the fine properties of the jump distribution, hence in particular we must specify the PDF of the jumps. A particularly relevant case is the Lévy stable family, with jumps having the characteristic function $\hat{\lambda}(k)=\exp(-|\gamma k|^\alpha)$, where $\gamma$ represents a length scale. Notably, although the characteristic function is known for each $\alpha$, the analytical form of the corresponding PDF is known only in few cases. This means that, contrary to the previous example, a complete exact study is out of reach. Nevertheless, several algorithms allow to generate jumps drawn from a symmetric stable law \cite{Kla-Sok}, hence a numerical investigation of the optimization problem is possible.

By considering the symmetric stable case, we are able to evaluate the values of $b$ for which Eq. \eqref{eq:Cond} is satisfied, and study how the range of allowed $b$ varies with $\alpha$. Since in this case Eq. $\eqref{eq:Cond}$ is sufficient, this allows us to identify regions of the $(\alpha,b)$ plane where a nontrivial optimization is guaranteed. We first observe that the result also depend on the value of the length scale $\gamma$, but we can remove this dependence by considering the rescaled parameter $b/\gamma$. The numerical evaluation of Eq. \eqref{eq:Cond} is conducted for various $\alpha$ in the range $0.7<\alpha<2$. For each value, we compute the survival probability after $1$ and $2$ steps as a function of $b/\gamma$. In every case, we find that Eq. \eqref{eq:Cond} is satisfied for $b/\gamma$ larger than a certain critical value, and remains valid beyond this value. Interestingly, the critical value becomes larger as $\alpha$ decreases. Below $\alpha=0.7$ it becomes difficult to compute the value of the critical distance, due to the excessive amount of statistics required to obtain results with sufficient accuracy. Nevertheless, we observe that the survival probabilities display a behaviour similar to the other cases, suggesting then that for any $0<\alpha<2$ there is a value of the initial distance from the target beyond which a nontrivial optimization of the MFPT exists, even though for small $\alpha$ this value may become very large.

To fix the ideas, let us consider the case $\alpha=1$, where $\lambda(\eta)$ has a simple analytical expression:
\begin{equation}
	\lambda(\eta)=\frac{1}{\gamma\pi}\frac{1}{1+(\eta/\gamma)^2}.
\end{equation}
We can thus write explicitly $Q_0(1;b)$ and $Q_0(2;b)$, and restate Eq. \eqref{eq:Cond} as
\begin{equation}
	\frac{\int_{0}^{\infty}\frac{\arctan(x)}{1+(\beta-x)^2}dx}{\pi+2\arctan(\beta)}<\frac12\arctan(\beta),
\end{equation}
where indeed $\beta=b/\gamma$. The lhs and rhs are both monotonic increasing functions of $\beta$ that converge to $\pi/4$. By studying this inequality numerically for values of $\beta$ up to $10^4$, we find that the lhs becomes smaller than the rhs for $\beta>\beta_c\approx1.8085$, and remains so as $\beta$ increases. This confirms our observation regarding the existence of a nontrivial optimization when the ratio $b/\gamma$ is larger than a critical value.

According to the results of our simulations, we may thus split the $(\alpha, b)$ plane into two distinct regions, I and II. The first region is where condition \eqref{eq:Cond} is satisfied, thus a nontrivial optimization is guaranteed. In the second region instead the same condition is not verified, which means that the MFPT is locally decreasing for $r\to1^-$. Even though a nontrivial optimization can not be excluded, let us observe that for $b=0$ we are sure that the MFPT is a monotonic decreasing function of $r$. Indeed, since the jumps are symmetric and continuous, a theorem of Sparre-Andersen \cite{Sparre} yields the generating function $\LT{Q}_0(z;0)=1/\sqrt{1-z}$, independently of the fine properties of the jumps. By plugging $\LT{Q}_0(z;0)$ into Eq. \eqref{eq:MFPT} one obtains
\begin{equation}
	\langle\tau_0(r)\rangle=\frac{1+\sqrt{r}}{\sqrt{r}},
\end{equation}
which is clearly monotonic in $(0,1]$ and attains its minimum at $r^*=1$. Hence for $b=0$ the optimization is trivial. By continuity, we expect that $\MFPT{r}$ remains a monotonic function in a neighbourhood of $b=0$, thus the optimization remains trivial when $b$ is not too large. Actually, our numerical experiments showed a monotonic decreasing MFPT for any $b$ in region II. We thus conjecture that in this domain the optimization is always trivial, and the minimal MFPT is obtained for $r^*=1$. Hence we interpret the phases in the diagram of Fig. \ref{fig:Levy_PD} as the regions where the optimization is nontrivial (I) and where it is instead obtained by resetting after each jump (II).

\section{Conclusions}\label{s:Conc}
In this paper, we investigated the optimization problem of the MFPT for jump processes with restart. We considered a resetting protocol that allowed us to study an intermediate situation between a random walk and an uncorrelated sequence of i.i.d. jumps all starting from the origin. We found a general condition for determining when restarting the process nontrivially reduces the MFPT with respect to the sequence of jumps. Remarkably, this condition only involves statistical properties of the random walk in absence of resetting after two jumps, therefore is easy to check numerically. When the process in absence of resetting has a MFPT larger than the sequence of jumps, the aforementioned condition is sufficient to establish the existence of a nontrivial resetting probability that minimizes the MFPT.

We have examined two relevant examples. For the skewed Laplace distribution, we drew the phase diagram in the space of a pair of parameters related to the bias of the jumps and the initial distance to the barrier. We found a rich situation where the system displays three distinct phases: a first phase, where a nontrivial optimization exists; a second phase, where the MFPT decreases monotonically with the resetting probability, so that the minimum is obtained for $r^*=1$; and a third phase, where the MFPT increases monotonically and $r^*=0$ yields the minimal MFPT. We note that this phase is observed only when the bias is directed toward the barrier. For symmetric Lévy flights with jumps drawn from Lévy stable laws, we drew the phase diagram in the space of Lévy indexes $\alpha$ and initial distances to the barrier $b$, showing the regions where such a nontrivial optimization exists. Remarkably, we found that for any $\alpha$ there is a critical value of $b$ beyond which the minimum is obtained for some $0<r^*<1$.

Before concluding, let us remark that the model of restart of Eq. \eqref{eq:dyn} can be extended to study more general situations or different problems than the one considered here. For example, consider
\begin{equation}
	x_{n}=\begin{cases}
		x_{n-1}+\eta_{n*}&\text{with Prob. }1-r\\
		\xi_{m(n)} &\text{with Prob. }r,
	\end{cases}
\end{equation}
where $m(n)$ is the number of restarts up to step $n$, $n^*=n-m(n)$ and $\xi_{i}$ are, for instance, Gaussian random variables with mean $x_0$. This resetting system can be used to model a scenario in which the relocation of the particle to $x_0$ at each restart event is influenced by Gaussian noise. This can simulate, for instance, the experimental error associated with returning the system to its initial position and aligns with an existing line of research exploring the impact of new resetting protocols, such as ``imperfect'' \cite{EvaMaj-2011-II,DahCheSch-2021,GonHidBar-2021,QueBoy-2023,MorOlsKri-2023,Ols-2023}, partial \cite{TalRoiReu-2022,DiBCheHar-2023} or ``smart'' resetting \cite{TalKeiReu-2024}. The study of such a model and other possible generalizations are left for future work.

\bibliographystyle{aip}

\begin{thebibliography}{127}
\bibitem [{Not()}]{Note}%
	\BibitemOpen
	\href@noop {} {}\bibinfo {note} {A more detailed analysis will follow in a
		separate paper.}\BibitemShut {Stop}%
\end{thebibliography}

\end{document}